\documentclass{iopart}
\def\<{{<}} 
\def\>{{>}}

\usepackage{graphicx}
\input{epsf}
\usepackage{amssymb}
\begin{document}

\title{Universal Emergence of PageRank} 

\author{K.M.Frahm$^1$, B.Georgeot$^2$ and D.L.Shepelyansky$^3$}
\address{Laboratoire de Physique Th\'eorique du CNRS, IRSAMC, 
Universit\'e de Toulouse, UPS, 31062 Toulouse, France}
\eads{\mailto{$^1$frahm@irsamc.ups-tlse.fr}, 
{$^2$georgeot@irsamc.ups-tlse.fr},
\mailto{$^3$dima@irsamc.ups-tlse.fr}}
% \date{March 25, 2011}
%\date{\today}

%\pacs{89.20.Hh, 89.75.Hc, 05.40.Fb, 64.60.F-}
%89.20.Hh       World Wide Web, Internet
%89.75.Hc       Networks and genealogical trees 
%05.40.Fb       Random walks and Levy flights
%72.15.Rn Localization effects (Anderson or weak localization) 
%87.23.Ge       Dynamics of social systems
% 64.60.F-      Equilibrium properties near critical points, critical exponents

\begin{abstract}
The PageRank algorithm enables to rank the nodes of a network
through a specific eigenvector of the Google matrix, using  
a damping parameter $\alpha \in ]0,1[$.  
Using  extensive
numerical simulations of large web networks, with a special accent on
British University networks, we determine
numerically and analytically the universal features of PageRank
vector at its emergence when $\alpha \rightarrow 1$. The whole network
can be divided into a core part and a group of invariant subspaces.
For $ \alpha \rightarrow 1$ the PageRank converges to a 
universal power law distribution on the invariant subspaces
whose size distribution also follows a universal power law.
The convergence of PageRank at  $ \alpha \rightarrow 1$
is controlled by  eigenvalues of the core part of the Google matrix
which are extremely close to unity leading to large relaxation 
times as for example in  spin glasses.
\end{abstract}

%\submitto{\NJP}
\maketitle
\section{Introduction}
The PageRank Algorithm (PRA) \cite{brin}
is a cornerstone element of the Google search engine
which allows to perform an efficient information
retrieval from the World Wide Web (WWW) and other
enormous directed networks created by the modern society
during  last two decades \cite{googlebook}.  
The ranking based on PRA finds applications in such diverse 
fields as Physical Review citation network
\cite{redner,fortunato}, scientific journals rating \cite{eigenfactor},
ranking of tennis players \cite{tennis} and many others
\cite{avrachenkov}.
The PRA allows to find efficiently 
the PageRank vector of the Google matrix 
of the network whose values
enable to rank the nodes.
For a given network with $N$ nodes the Google matrix 
is defined as
\begin{equation}
   \mathbf{G} = \alpha  \mathbf{S} + (1-\alpha) 
 e e^T/N \;\; ,
\label{eq1} 
\end{equation} 
where the matrix $\mathbf{S}$ is obtained from an adjacency matrix 
$\mathbf{A}$ by normalizing all nonzero colummns to one ($\sum_j S_{ij}=1$) 
and replacing columns with only zero elements by $1/N$ ({\em dangling nodes}).
For the WWW an element $A_{ij}$ of the adjacency matrix 
is equal to unity
if a node $j$ points to node $i$ and zero otherwise.
Here $e=(1,\ldots,1)^T$
is the unit column vector and 
$e^T$ is its transposition.
The damping parameter $\alpha$ in the WWW context 
describes the probability 
$(1-\alpha)$ to jump to any node for a random surfer. 
For WWW the Google search uses $\alpha \approx 0.85$  
\cite{googlebook}.

The matrix $\mathbf{G}$ belongs to the class 
of Perron-Frobenius operators 
naturally appearing for  Markov chains
and dynamical systems \cite{googlebook,mbrin}.
For $0< \alpha <1$ there is only one maximal eigenvalue 
$\lambda=1$ of $\mathbf{G}$. The corresponding eigenvector
is the PageRank vector which has nonnegative components $P(i)$
with $\sum_{i}P(i)=1$,
which can be ranked in decreasing order to give the
PageRank index $K(i)$. For WWW it is known that
the probability distribution $w(P)$ of $P(i)$ values
is described by a power law 
$w(P) \propto 1/P^{\mu}$ 
with $\mu \approx 2.1$ \cite{donato},
corresponding to the related cumulative dependence
$P(i) \propto 1/K^\beta(i)$ with $\beta=1/(\mu-1) \approx 0.9$
at $\alpha \sim 0.85$.

\begin{figure}[t]
\begin{indented}\item[]
\begin{center}  
\includegraphics[width=0.7\textwidth]{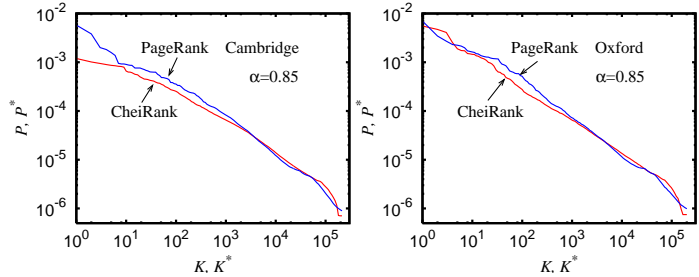}
% \vglue -0.3cm
\caption{
PageRank $P$  and CheiRank $P^*$  
versus the corresponding rank indexes $K$ and $K^*$ for the WWW
networks of Cambridge 2006 (left panel) and Oxford 2006 (right panel);
here $N=212710$ ($200823$) and the number of 
links is $L=2015265$ ($1831542$) for Cambridge (Oxford).}
\label{fig1}
\end{center}
\end{indented}
\end{figure}

The PageRank performs ranking which in average is 
proportional to the number of ingoing links \cite{googlebook,litvak},
putting at the top the most known and popular nodes.
However, in certain networks outgoing links 
also play an important role.
Recently, on the examples  of the procedure call network of
Linux Kernel software \cite{linux}
and the Wikipedia articles network \cite{wiki}, it was shown that 
a relevant additional ranking is obtained by    
considering the network
with inverse link directions in the adjacency matrix
corresponding to $(A_{ij}) \rightarrow \mathbf{A}^T=(A_{ji})$
and constructing from it a reverse Google matrix $\mathbf{G^*}$
according to relation (\ref{eq1}) at the same $\alpha$.
The eigenvector of $\mathbf{G^*}$ with eigenvalue 
$\lambda=1$ gives then a new  PageRank $P^*(i)$
with ranking index $K^*(i)$, which was named 
CheiRank \cite{wiki}.
It rates nodes in average proportionally to 
the number of outgoing links highlighting their communicative
properties \cite{linux,wiki}. For WWW one finds $\mu \approx 2.7$
\cite{donato} so that the decay of CheiRank $P^* \propto 1/{K^*}^\beta$
is characterized by a slower decay exponent $\beta \approx 0.6$
compared to PageRank. In Fig.~\ref{fig1}, we show PageRank and CheiRank 
distributions for the WWW networks of the 
Universities of Cambridge and Oxford  
(2006), obtained from the database \cite{uniuk}.

Due to importance of PageRank for information retrieval
and ranking of various directed networks \cite{avrachenkov}
it is important to
understand how it is affected by the variation of the damping parameter
$\alpha$. In the limit $\alpha\to 1$ the PageRank is determined 
by the eigenvectors 
of the highly degenerate eigenvalue $1$ \cite{capizzano}. These eigenvectors
correspond by definition to invariant subspaces through the matrix 
$\mathbf{S}$.  
It is known
\cite{vigna} that in general
these subspaces correspond
to sets of nodes with ingoing links from the rest of the network
but no outgoing link to it.  These parts of the network
have been given different names in the literature (rank sink, 
out component, bucket, and so on). In this paper, we show that 
for large matrices of size up to several millions
the structure of these
invariant subspaces is universal and study  
in detail the universal behavior of the PageRank at $\alpha\to 1$
related to the spectrum of  $\mathbf{G}$, 
using an optimized Arnoldi algorithm.

We note that this behavior is linked to the internal structure of the network. 
Indeed, it is possible to randomize real networks by randomly exchanging the 
links while keeping exactly the same number of ingoing and 
outgoing links. It was shown in \cite{giraud} that this process 
generally destroys the structure of the network and creates a huge 
gap between the first unit eigenvalue and the second eigenvalue (with 
modulus below $0.5$). In this case the PageRank simply 
goes for $\alpha\to 1$ to the unique eigenvector of the matrix 
$\mathbf{S}$ associated with the unit eigenvalue.

The paper is organized as follows: in Section 2 we
discuss the spectrum and subspace structure of the Google matrix;
in Section 3 we present the construction of invariant subspaces,
the numerical method of PageRank computation at small damping factors
is given in Section 4, the projected power method is described in Section 5,
universal properties of PageRank are analyzed in Section 6
and discussion of the results is given in Section 7. 

\section{Spectrum and subspaces of Google matrix}

In order to obtain the invariant subspaces, 
for each node we determine iteratively the set
of nodes that can be reached 
 by a chain of non-zero matrix elements. 
If this set contains all nodes of the network, 
we say that the initial node belongs to the {\em core space} $V_c$. 
Otherwise, the limit set defines a subspace which 
is invariant with respect to applications of the matrix $\mathbf{S}$. 
In a second step we merge all subspaces with common 
members, and obtain a sequence of disjoint subspaces $V_j$ of dimension 
$d_j$ invariant by applications of $\mathbf{S}$.
This scheme, which can be efficiently implemented in a 
computer program, provides a subdivision of network nodes  in $N_c$ 
core space nodes (typically 70-80\% of $N$) and $N_s$ subspace nodes 
belonging to at least one of the invariant subspaces $V_j$ 
inducing the block triangular structure, 
\begin{equation}
\label{Smat_block}
\mathbf{S}=\left(\begin{array}{cc}
\mathbf{S_{ss}} & \mathbf{S_{sc}}  \\
0 & \mathbf{S_{cc}}  \\
\end{array}\right)
\end{equation}
where the subspace-subspace block $\mathbf{S_{ss}}$ is actually composed of 
many diagonal blocks for each of the invariant subspaces. Each of these blocks 
correspond to a column sum normalized matrix of the same type as $\mathbf{G}$ 
and has therefore at least one unit eigenvalue thus explaining the 
high degeneracy. Its eigenvalues and eigenvectors are easily accessible by 
numerical diagonalization (for full matrices) thus allowing to count the 
number of unit eigenvalues, e.g. 1832 (2360) for the WWW
networks of Cambridge 2006 (Oxford 2006) and also to verify that all 
eigenvectors of the unit eigenvalue are in one of the subspaces. 
The remaining eigenvalues of $\mathbf{S}$ can be obtained from 
the projected core block 
$\mathbf{S_{cc}}$ which is not column sum normalized 
(due to non-zero matrix elements 
in the block $\mathbf{S_{sc}}$) and has therefore eigenvalues 
strictly inside the 
unit circle $|\lambda^{\rm (core)}_j|<1$. We have applied the 
Arnoldi method (AM) 
\cite{arnoldibook,golub,ulamfrahm} with Arnoldi dimension $n_A=20000$ 
to determine the largest eigenvalues of $\mathbf{S_{cc}}$. 
For both example networks 
this provides at least about 4000 numerical accurate eigenvalues in the range 
$|\lambda|\ge 0.7$. For the two networks 
the largest core space eigenvalues are given by 
 $\lambda^{\rm (core)}_1=0.999874353718$ (0.999982435081) 
with a quite clear gap 
$1-\lambda^{\rm (core)}_1 \sim 10^{-4}$ ($\sim 10^{-5}$). 
We also mention that the largest subspace eigenvalues 
with modulus below 1 also have a comparable gap $\sim 10^{-5}$. 
In order to obtain this accuracy it is highly important 
to apply the AM to $\mathbf{S_{cc}}$ and not to the full matrix 
$\mathbf{S}$ (see more details below). 
In the latter 
case the AM fails to determine the degeneracy of the unit eigenvalue 
and for the 
same value of $n_A$ it produces less accurate results. 
\begin{figure}[t]
\begin{indented}\item[]
\begin{center}  
\includegraphics[width=0.7\textwidth]{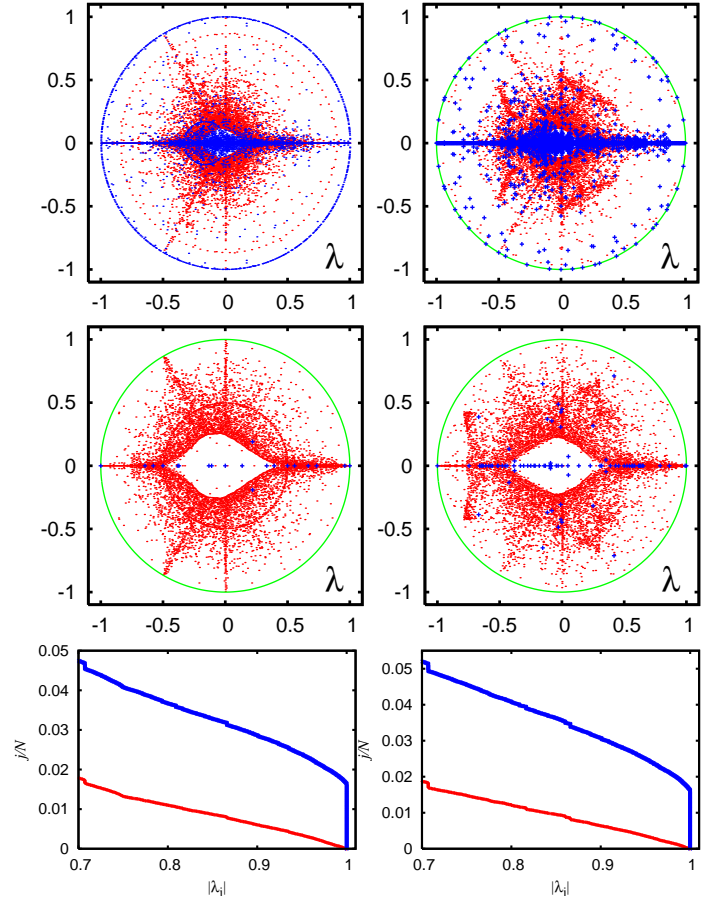}
% \vglue -0.3cm
\caption{Left panels (right panels) 
correspond to Cambridge 2006 (Oxford 2006).
{\em Top row:} Subspace eigenvalues of the matrix $\mathbf{S}$ 
(blue dots or crosses) and 
core space eigenvalues (red dots) in $\lambda-$plane 
(green curve shows unit circle); 
here $N_s=48239$ (30579),
There are 1543 (1889)
invariant subspaces, with maximal dimension 4656 
(1545) and 
the sum of all subspace dimensions is 
$N_s=48239$ (30579). 
The core space eigenvalues are obtained from the Arnoldi method applied to 
the block $S_{cc}$ with 
Arnoldi dimension 20000 and are numerically accurate for $|\lambda|\ge 0.7$.
{\em Middle row:} Eigenvalue spectrum for the matrix $\mathbf{S}^*$,
corresponding to the CheiRank, for
Cambridge 2006 (left panel) and Oxford 2006 (right panel) 
with red dots for core space eigenvalues (obtained 
by the Arnoldi method applied to $\mathbf{S_{cc}}^*$ with $n_A=15000$), 
blue crosses for 
subspace eigenvalues and the green curve showing the unit circle.
{\em Bottom row:} 
Fraction $j/N$ of eigenvalues with $|\lambda| >  |\lambda_j|$ for the 
core space eigenvalues (red bottom curve) and all eigenvalues (blue top curve)
from top row data. 
The number of 
eigenvalues with $|\lambda_j|=1$ is 3508 (3275) of which 
1832 (2360) are at $\lambda_j=1$;
it larger than the number of invariant subspaces which 
have each at least one unit eigenvalue.}
\label{fig2}
\end{center}
\end{indented}
\end{figure}
 
In Fig. \ref{fig2} we present the spectra of 
subspace and core space eigenvalues in the complex plane 
$\lambda$ as well as 
the fraction of eigenvalues with modulus larger than $|\lambda|$, showing
that subspace eigenvalues are spread around the unit circle 
being closer to $|\lambda|=1$ than core eigenvalues.
The fraction of states with $|\lambda| >  |\lambda_j|$ has
a sharp jump at $\lambda=1$, corresponding to the contribution of $N_s$,
followed by an approximate linear growth.

We now turn to the implications of this structure to the 
PageRank vector $P$; it can be formally expressed as 
\begin{equation}
\label{PageRank1}
P=(1-\alpha)\,(\mathbf{1}-\alpha\mathbf{S})^{-1}\,e/N .
\end{equation}
Let us first assume that $\mathbf{S}$ is diagonalizable (with no non-trivial Jordan 
blocks). We denote by $\psi_j$ its (right) eigenvectors and expand the 
vector $N^{-1}\,e=\sum_j c_j\,\psi_j$ in this eigenvector basis with 
coefficients $c_j$. Inserting this expansion in Eq. (\ref{PageRank1}), 
we obtain
\begin{equation}
\label{PageRank2}
P=\sum_{\lambda_j=1} c_j\,\psi_j + \sum_{\lambda_j\neq 1} 
\frac{1-\alpha}{(1-\alpha)+\alpha(1-\lambda_j)}\, c_j\,\psi_j \ .
\end{equation}
In the case of non-trivial Jordan blocks we may have in the second sum 
contributions $\sim (1-\alpha)/(1-\alpha\,\lambda_j)^q$ with some 
integer $q$ smaller or equal to the size of the Jordan block \cite{capizzano}. 
Suppose we have for example a Jordan block of dimension 2 with a principal 
vector $\tilde \psi_j$ such that 
$\mathbf{S}\,\tilde \psi_j=\lambda_j\tilde \psi_j+\psi_j$ with $\psi_j$ the 
corresponding eigenvector. From this we obtain for arbitrary integer $n$ 
the following condition on the 1-norm of these vectors~: 
$\|\tilde \psi_j\|_1\ge 
\|\mathbf{S}^n \tilde \psi_j\|_1=
\|\lambda_j^n\tilde \psi_j+n\lambda_j^{n-1}\psi_j\|_1
\ge \Bigl||\lambda_j|^n\|\tilde \psi_j\|_1-n|\lambda_j|^{n-1}
\|\psi_j\|_1\Bigr|$ showing that one should have  
$\psi_j = 0$ if $|\lambda_j|=1$. Even if 
$|\lambda_j|<1$  this condition is hard to fulfill
for all $n$ if $|\lambda_j|$ is close to 1. In general 
the largest eigenvalues with modulus below 1 are not likely to 
belong to a non-trivial Jordan block; this is indeed well verified for 
our university networks since the 
largest core space eigenvalues are not degenerate. 

Here Eq. (\ref{PageRank2}) indicates that 
in the limit $\alpha\to 1$ the PageRank
converges to a particular linear combination of the eigenvectors 
with $\lambda= 1$, which 
are all localized in one of the subspaces. For a finite value of $1-\alpha$ 
the scale of this convergence is set by the condition 
$1-\alpha\ll 1-\lambda_1^{\rm (core)}\sim 10^{-4}$ 
($10^{-5}$) and the corrections for 
the contributions of the core space nodes are 
$\sim (1-\alpha)/(1-\lambda_1^{\rm (core)})$. In order to test 
this behavior we have numerically computed the PageRank vector for 
values $10^{-8}\le 1-\alpha\le 0.15$.  For $1-\alpha\approx10^{-8}$,
the usual power method (iterating the matrix $\mathbf{G}$ on an initial vector)
is very slow and in many cases fails to converge with a reasonable precision. 
In order to get the PageRank vector in this regime, 
we use a combination of power
and Arnoldi methods that allowed us to reach the precision
$\|P-\mathbf{G}(\alpha)P\|_1<10^{-13}$:
after each $n_i$ iterations with the power method 
we use the resulting vector as initial vector for an Arnoldi diagonalization 
choosing an Arnoldi matrix size $n_A$; the resulting eigenvector for the 
largest eigenvalue is used as a new vector to which we apply the power 
method and so on until convergence by the condition 
$\|P-\mathbf{G}(\alpha)P\|_1<10^{-13}$ is reached. 
For the university 
network data of \cite{uniuk} in most cases the values $n_i=10^4$ and 
$n_A=100$ ($n_A=500$ for Cambridge 2006) provide convergence 
with about $\sim 10$ iterations of the process
(for $1-\alpha=10^{-8}$). Additional details are given below.

\section{Construction of invariant subspaces}

In order to construct the invariant subspaces we use the following scheme 
which we implemented in an efficient computer program. 

For each node $j=1,\,\ldots,\,N$ we determine iteratively a sequence of 
sets $E_n$, with $E_0=\{j\}$ and $E_{n+1}$ containing the nodes $k$ 
which can be reached by a non-zero matrix element $S_{kl}$ from 
one of the nodes $l\in E_n$. 
Depending on the initial node $j$ there are two possibilities:  
a) $E_n$ increases with the iterations until it contains all nodes of 
the network, especially if one set $E_n$ contains a dangling node connected 
(by construction of $\mathbf{S}$) to all other nodes, or 
b) $E_n$ saturates at a limit 
set $E_{\infty}$ of small or modest size $d_j<N$. In the first case, 
we say that the node $j$ belongs to the {\em core space} $V_c$. 
In the second case the limit set defines a subspace $V_j$ of dimension 
$d_j$ which is invariant with respect to applications of the matrix 
$\mathbf{S}$. 
We call the initial node $j$ the {\em root node} of this subspace;
the members of $E_{\infty}$ do not need to be tested themselves as 
initial nodes subsequently since they are already identified as 
{\em subspace nodes}. 
If during the iterations a former root node appears as a member in a new 
subspace one can absorb its subspace in the new one and this node loses 
its status as root node. Furthermore, the scheme is greatly simplified if 
during the iterations a dangling node or another node already being 
identified as core space node is reached. In this case 
one can immediately attribute 
the initial node $j$ to the core space as well. 

For practical reasons it may be useful to stop the iteration if the set 
$E_n$ contains a macroscopic number of nodes larger than $B\,N$ where $B$ is 
some constant of order one and to attribute in this case the node $j$ to 
the core space. This does not change the results provided that 
$B\,N$ is above the maximal subspace dimensions. For the university networks 
we studied,
the choice $B \ge 0.1$ turned out to be sufficient since there is always a 
considerable number of dangling nodes. 

In this way, we obtain a subdivision of the nodes of the network in $N_c$ 
core space nodes (typically 70-80\% of $N$) and  $N_s$ subspace nodes 
belonging to at least one of the invariant subspaces $V_j$. However, at this 
point it is still possible, even likely, that two subspaces have common 
members. Therefore in a second step we merge all subspace with common 
members and choose arbitrarily one of the root nodes 
as the ``root node'' of the new bigger subspace which is of course also 
invariant with respect to $\mathbf{S}$.

We can also mention that most of the subspaces contain one or more 
``zero nodes'' (of first order) with outgoing links to the subspace but no 
incoming links from the same or other subspaces (but they may have incoming 
links from core space nodes as every subspace node). 
These nodes correspond to complete zero lines in the corresponding diagonal 
block for this subspace in the matrix $\mathbf{S}$ and 
therefore they produce a trivial eigenvalue zero. 
Furthermore, there are also zero nodes of higher order $j$ ($\ge 2$) 
which have incoming subspace links only from other zero nodes of order $j-1$ 
resulting in a non-trivial Jordan block structure with eigenvalue zero. 
In other words, when one applies the matrix $\mathbf{S}$ 
to a vector with non-zero 
elements on all nodes of one subspace one eliminates successively the zero 
nodes of order $1,\,2,\,3,\,\ldots$ and finally the resulting 
vector will have non-zero values only for the other  ``non-zero nodes''. 
Due to this any subspace eigenvector of $\mathbf{S}$ 
with an eigenvalue different 
from zero (and in particular the PageRank vector) cannot have any contribution 
from a zero node. 

In a third step of our scheme we therefore determined the 
zero nodes (of all orders) and the reduced subspaces without these zero nodes. 
The results for the distribution of subspace dimensions 
is discussed in Section 6 
(see  the left panel of Fig.~\ref{fig7}).
The distribution is 
essentially unchanged if we use the reduced 
subspaces since the number of zero nodes is
below $10\%$ of $N_s$ for most of universities. Only for the matrix 
$\mathbf{S}^*$ of Wikipedia we have about 
$45\%$ of zero nodes that reduces the value of $N_s$ 
from 21198 to 11625.

Once the invariant subspaces of $\mathbf{S}$ are known 
it is quite obvious to obtain 
numerically the exact eigenvalues of the subspaces, including the exact 
degeneracies. Thus, using the Arnoldi method we determine 
the largest remaining eigenvalues of the core 
projected block $\mathbf{S_{cc}}$. 
In Fig.~\ref{fig2} the complex spectra of subspace and core space 
eigenvalues of  $\mathbf{S}$ and  $\mathbf{S}^*$ are shown for the two 
networks of Cambridge 2006 and Oxford 2006 as well as the fraction 
of eigenvalues with modulus larger than $|\lambda|$ indicating a 
macroscopic fraction of about 2\% of eigenvalues with $|\lambda_j|=1$. 

In Table 1, we summarize the main quantities of 
networks studied: network size $N$, 
number of network links $L$, number of subspace nodes 
$N_s$ and average subspace dimension $\langle d\rangle$ for the 
university networks considered in Fig.~\ref{fig4} and the matrix $S^*$ 
of Wikipedia. 
%\bigskip

\begin{table}[h]%
\small
\caption{\label{table1} Network parameters}
\begin{tabular}{|l|c|c|c|c|}
\hline
 & $N$ & $L$ & $N_s$ & $\langle d\rangle$ \\
\hline
\hline
Cambridge 2002 & 140256 & 752459 & 23903 & 20.36 \\
\hline
Cambridge 2003 &\ 201250\ \ &\ 1182527\ \ &\ 45495\ \ &\ 24.97\ \ \\
\hline
Cambridge 2004 & 206998 & 1475945 & 44181 & 26.14 \\
\hline
Cambridge 2005 & 204760 & 1505621 & 44978 & 29.30 \\
\hline
Cambridge 2006 & 212710 & 2015265 & 48239 & 31.26 \\
\hline
Oxford 2002 & 127450 & 789090 & 14820 & 14.01 \\
\hline
Oxford 2003 & 144783 & 883672 & 19972 & 19.85 \\
\hline
Oxford 2004 & 162394 & 1158829 & 29729 & 19.18 \\
\hline
Oxford 2005 & 169561 & 1351932 & 36014 & 23.34 \\
\hline
Oxford 2006 & 200823 & 1831542 & 30579 & 16.19 \\
\hline
Glasgow 2006 & 90218 & 544774 & 20690 & 28.54 \\
\hline
Edinburgh 2006 & 142707 & 1165331 & 24276 & 26.24 \\
\hline
UCL 2006 & 128450 & 1397261 & 25634 & 28.64 \\
\hline
Manchester 2006 & 99930 & 1254939 & 23648 & 26.07 \\
\hline
Leeds 2006 & 94027 & 862109 & 12605 & 31.20 \\
\hline
Bristol 2006 & 92262 & 1004175 & 9143 & 19.49 \\
\hline
Birkbeck 2006 & 54938 & 1186854 & 3974 & 19.11 \\
\hline
Wikipedia ($S^*$) & 3282257 & 71012307 & 21198 & 3.96 \\
\hline
\end{tabular}
\end{table}

\section{Numerical method of PageRank computation}
Let us now discuss the numerical techniques that we developed
in order to compute the PageRank.
The standard method to determine the PageRank is the power method
\cite{brin,googlebook}. However, 
this method fails to converge at a sufficient rate in the limit $\alpha\to 1$ 
and therefore we need a more refined method. First we briefly discuss how the 
power method works and then how it can be modified to improve the convergence.

Let $P_0$ be an initial vector which is more or less a good approximation 
of the PageRank. Typically one may choose $P_0=e/N$ where 
$e=(1,\,\ldots,\,1)^T$. For simplicity let us also suppose that the 
matrix $\mathbf{G}(\alpha)$ can be diagonalized. The eventual existence 
of principal vectors and non-trivial Jordan blocks does not change the 
essential argument and creates only minor technical complications. 
The initial vector can be developed in the eigenvector 
basis of $\mathbf{G}(\alpha)$ as:
\begin{equation}
\label{eqPhi0}
P_0=P+\sum_{j\ge 2} C_j\,\varphi_j
\end{equation}
where $P=\varphi_1$ is the exact PageRank, which is for $\alpha<1$ the only 
(right) eigenvector of $\mathbf{G}(\alpha)$ with eigenvalue 1. 
Here $\varphi_j$ denote for $j\ge 2$ 
other (right) eigenvectors with eigenvalues $\lambda_j$ 
such that $|\lambda_j|\le \alpha$ and $C_j$ are the expansion coefficients. 
We note that $e^T\varphi_j=0$ for $j\ge 2$ since $e$ is 
the first left eigenvector bi-orthogonal to other right eigenvectors 
and for sufficiently small $C_j$ the expansion coefficient of $P$ in 
$P_0$ is exactly 1 if $P_0$ and $P$ are both normalized by the 1-norm. 
Iterating the initial vector by $\mathbf{G}(\alpha)$ one obtains after
$i$ iterations~:
\begin{equation}
\label{eqPhii}
P_i=\mathbf{G}^i(\alpha)\,P_0
=P+\sum_{j\ge 2} C_j\,\lambda_j^i\,\varphi_j\ .
\end{equation}
Therefore the convergence of the power method goes with $\sim\lambda_2^i$ 
where $\lambda_2$ is the second largest eigenvalue. In the case of realistic 
networks $\lambda_2$ is typically 
highly degenerate and equal to $\alpha$. Typically 
there are also complex eigenvalues with non-trivial phases  where only the 
modulus is equal to $\alpha$ and whose contributions imply the same speed of 
convergence. In the limit $\alpha\to 1$ the power method becomes highly 
ineffective due to these eigenvalues. 
For example to verify the condition $\alpha^i<\varepsilon$ one needs 
$i>3\cdot 10^9$ iterations for $1-\alpha=10^{-8}$ and 
$\varepsilon=10^{-13}$. 

\begin{figure}[t]
\begin{indented}\item[]
\begin{center}  
\includegraphics[width=0.7\textwidth]{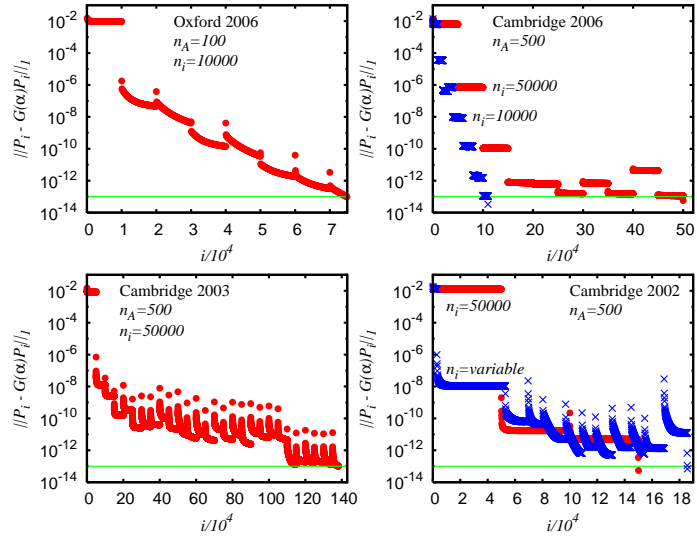}
% \vglue -0.3cm
\caption{Convergence of the combined power-Arnoldi method to calculate the 
PageRank for $1-\alpha=10^{-8}$. Shown is the quantity 
$\|P_i-\mathbf{G}(\alpha)P_i\|_1$ to characterize the quality of the 
approximate PageRank $P_i$ 
versus the number of iterations $i$ done 
by the power method. The green line at $10^{-13}$ shows the line below 
which convergence is reached. 
The upper left panel shows the data for Oxford 2006 with $n_A=100$ 
and $n_i=10000$. The upper right panel corresponds to 
Cambridge 2006 with $n_A=500$ and $n_i=50000$ (red dots) or 
$n_i=10000$ (blue crosses). 
The lower left panel shows the case Cambridge 2003 with $n_A=500$ and 
$n_i=50000$ for which it is particularly hard to obtain convergence. 
The lower right panel compares for the case Cambridge 2002 the 
choice $n_A=500$ and $n_i=50000$ (red dots) with $n_A=500$ and 
$n_i=$ variable (blue crosses) with $n_i$ determined by the criterion 
that the relative change of $\|P_i-\mathbf{G}(\alpha)P_i\|_1$ between $i$ and 
$i+100$ is less than $10^{-4}$.
}
\label{fig3}
\end{center}
\end{indented}
\end{figure}

In order to obtain a faster convergence we propose a different method based 
on the Arnoldi method 
\cite{arnoldibook,golub,ulamfrahm}. The idea of the Arnold method is to 
diagonalize the matrix representation of $\mathbf{G}(\alpha)$ on the Krylov 
space generated by $P_0,\,P_1,\,\ldots,\,P_{n_A-1}$ where we call 
$n_A$ the Arnoldi dimension. For reasons of numerical stability one 
constructs by Gram-Schmidt orthogonalization an orthogonal basis of the 
Krylov space which also provides the matrix elements of the 
matrix representation of $\mathbf{G}(\alpha)$ in this basis. 
In the particular case where the number of 
non-vanishing coefficients $C_j$ in Eq.~(\ref{eqPhi0}) is not too large the 
Arnoldi method should even provide the exact PageRank, obtained 
as the eigenvector of the largest eigenvalue on the Krylov space, and 
exactly suppress 
the other eigenvector contributions provided that the dimension $n_A$ 
of the Krylov space is sufficiently large to contain all other eigenvectors 
contributing in Eq.~(\ref{eqPhi0}). Of course in reality the number of 
non-vanishing coefficients $C_j$ is not small but one can use a strategy 
which consists first to apply the power method with 
$n_i$ iterations to reduce the contributions of the big majority of 
eigenvectors whose eigenvalues have a reasonable gap from the unit circle and 
in a second step the Arnoldi method to eliminate the remaining 
``hard'' eigenvectors whose eigenvalues 
are too close to the unit circle for the power method. 
Even though this strategy does not provide the numerical 
``exact'' PageRank, it considerably improves the quality of the initial 
vector as approximation of the PageRank and repeating this scheme on the 
new approximation as initial vector  
(with suitable values for $n_i$ and $n_A$) one obtains an algorithm 
which efficiently computes the PageRank up to a high precision as 
can be seen in Fig.~\ref{fig3}. To measure the quality of the 
PageRank vector we compute the quantity 
$\|P_i-\mathbf{G}(\alpha)P_i\|_1$ and iterate our algorithm until 
this quantity is below $10^{-13}$. Using this convergence criterion 
for most university networks from the database \cite{uniuk} the 
choice of $n_i=10000$ and $n_A=100$ provides convergence with typically 
about $10$ steps of this procedure.

In Fig.~\ref{fig3} we show the convergence of this method for 
several university network cases with the initial vector 
$P_0=e/N$ and $1-\alpha=10^{-8}$. 
The typical situation is shown in the upper left 
panel for Oxford 2006. During the first power method 
cycle there is nearly no improvement of the quality of the PageRank.
This is completely 
normal in view of the small value of $1-\alpha$. However, the first Arnoldi 
step improves the quality by 4 orders of magnitude. Then the subsequent 
power method iterations of the second cycle continue to improve the 
convergence quality but their effect 
saturates after a certain number of iterations. 
The second Arnoldi step seems at first to reduce 
the PageRank quality but 
after a few number of power method iterations (in the third cycle) this loss 
is compensated and its quality improves until the next saturation and the 
next Arnoldi step. In total this provides a nice exponential convergence 
and after 7 Arnoldi steps and 75000 power method iterations in total 
the convergence is reached with very high accuracy. 
Apparently the Arnoldi method is rather efficient to reduce the coefficients 
$C_j$ associated to the eigenvectors with eigenvalues close to the circle of 
radius $\alpha$ but the approximation due to 
truncation of the Arnoldi matrix to the 
Krylov space at $n_A$ creates some 
artificial contributions from other eigenvectors whose eigenvalues have 
a quite big gap from 1 and whose contributions may be eliminated 
by a relatively 
modest number of power method iterations. 

The number $n_A=100$ appears very modest if compared to the degeneracy 
of the second eigenvalue $\lambda_2=\alpha$ which may easily be about 
1000-2000. Fortunately, the exact degeneracy of the eigenvalues close to or 
on the circle of radius $\alpha$ does not really count, since for each 
degenerate 
eigenspace only {\em one} particular eigenvector appears in the expansions 
(\ref{eqPhi0}), (\ref{eqPhii}) which can be relatively easily ``eliminated'' 
by an Arnoldi step with modest value of $n_A$. However, the total number 
of {\em different} eigenvalues (with different phases) on the circle of 
radius $\alpha$ is important and if this number is too big the convergence of 
the method is more difficult. This is actually the case for the university 
networks of Cambridge as can be seen in the upper left panel of 
Fig.~\ref{fig2} where the subspace eigenvalues of $\mathbf{S}$ 
for Cambridge 2006 
nearly fill out the unit circle and indeed we have to increase for 
these cases the Arnoldi dimension to $n_A=500$ in order to achieve a 
reasonable convergence. In the upper right panel of Fig.~\ref{fig3} we show 
the PageRank convergence for Cambridge 2006 with $n_A=500$ and two choices of 
$n_i=10000$ and $n_i=50000$. For this particular example the first choice is 
more efficient but this is not systematic and is different 
for other cases. We also see that increasing the value of $n_i$ the 
convergence is not immediately improved (the PageRank error does not 
really decrease during the power method cycle) but the positive effect of 
the next Arnoldi step 
will be much better, apparently because the bigger number of power method 
iterations allows to reduce the effect of more eigenvectors in the 
eigenvector expansion of $P_i$. 
In the lower left panel of Fig.~\ref{fig3} we show the case of Cambridge 2003 
which is particularly hard for the convergence and requires 28 Arnoldi steps 
with $n_i=50000$ and $n_A=500$. Actually here the choice $n_i=10000$ (not 
shown in the figure) is less efficient with nearly the doubled number of 
power method iterations and about 235 Arnoldi steps. 
In the lower right panel we consider the 
case of Cambridge 2002 where we need 3 Arnoldi steps for the parameters 
$n_A=500$ and $n_i=50000$. For this case, we also tried a different strategy 
which consists of using a variable value of $n_i$ determined by the criterion 
that when the relative change of $\|P_i-\mathbf{G}(\alpha)P_i\|_1$ from $i$ to 
$i+100$ is below $10^{-4}$ we perform one Arnoldi step but at latest 
after $50000$ power method iterations for each cycle. 
For this example this strategy does not really pay off since the 
overall number of power method iterations is even slightly increased 
and additionally we have 11 instead of 3 quite expensive Arnoldi steps. 
However, this approach 
has the advantage that one does not need to search in advance which exact 
choice of $n_i$ parameters works best. 
In practical calculations when calculating the PageRank for a continuous set 
of values of $\alpha$ one may also improve convergence simply by using the 
PageRank at a certain value of $\alpha$ as initial vector for the next 
value $\alpha+\Delta\alpha$. However, in Fig.~\ref{fig3}, we simply used the 
same initial vector $P_0=e/N$ for all cases 
in order to study the effectiveness of the method as such. 

The computational costs of the method are increased 
quite strongly with $n_A$ since 
the Arnoldi steps correspond to $n_A^2\,N+n_A\,L$ 
elementary operations (with $L$ being the number of links in the network) 
due to the Gram-Schmidt orthogonalization scheme and $n_A$ applications 
of $\mathbf{G}(\alpha)$ on a vector while one step 
with the power method costs $L$ operations. Therefore one Arnoldi step 
corresponds to $\sim (n_A^2\,(N/L) + n_A)$ steps of the power method which 
is $\sim 1000$ ($\sim 25000$) for $n_A=100$ ($n_A=500$) and $L/N\sim 10$ 
(typical value for most university networks of \cite{uniuk}). 

We mention that the method does not converge if we use only Arnodi steps 
without intermediate power method iterations (i.~e. $n_i=0$). 
Golub {\it et al.} 
\cite{golub} have suggested a different variant of the Arnoldi method 
where they determine the improved vector not as the eigenvector of the 
largest eigenvalue 
of the truncated squared Arnoldi matrix but as the vector corresponding to 
the smallest singular value of a matrix obtained from the full non-truncated 
rectangular Arnoldi matrix. We have also implemented this variant and we 
have confirmed for some examples that convergence by simply repeating 
these ``refined'' 
Arnoldi steps is possible but in general the computational time for 
convergence is much longer if compared to our method. We have also tested 
the combination of power method and refined Arnoldi steps and we find that 
this approach is in general comparable to our first method with a slight 
advantage for one or the other method depending on the network that is studied.

\section{Projected power method for the case of small core space 
eigenvalue gap}

The behavior of the PageRank in the limit $\alpha\to 1$ is 
determined by the core space eigenvalue gap $1-\lambda_1^{\rm (core)}$ 
where $\lambda_1^{\rm (core)}<1$ is the maximal eigenvalue 
of the core space projected matrix $\mathbf{S_{cc}}$ 
[see Eq.~(\ref{Smat_block})]. This eigenvalue and its 
eigenvector $\psi_1^{\rm (core)}$ can in principle be determined by the 
Arnoldi method applied to 
$\mathbf{S_{cc}}$. However, for certain university networks of \cite{uniuk}, 
Cambridge 2002, 2003, 2005 and Leeds 2006, we find that 
$\lambda_1^{\rm (core)}$ is extremely close to 1. Since the results 
of the Arnoldi method are obtained by standard double precision arithmetic
operations 
it gives a largest core space eigenvalue which is 
{\em numerically} equal to 1 for these cases (up to an error of order 
$\sim 10^{-14}$), This is not sufficient to provide 
an accurate value for the gap $1-\lambda_1^{\rm (core)}$ apart 
from the information that this gap is below $10^{-14}$.

To overcome this computational problem
we note that $\lambda_1^{\rm (core)}$ and $\psi_1^{\rm (core)}$ 
can also be numerically determined by a different algorithm. 
The main idea is to apply the power method, 
eventually with intermediate Arnoldi steps to accelerate convergence, as 
described in the previous section, 
to the matrix $\mathbf{S_{cc}}$ which 
first provides the eigenvector $\psi_1^{\rm (core)}$ and once the eigenvector 
is known its eigenvalue is simply obtained as 
$\lambda_1^{\rm (core)}=\|\mathbf{S_{cc}}\,\psi_1^{\rm (core)}\|_1$
if the normalization is given by $\|\psi_1^{\rm (core)}\|_1=1$. 
In this Section it is understood that $\mathbf{S_{cc}}$ 
is the matrix $\mathbf{S}$ multiplied left and right by the projection 
operator on the core space (and similarly for $\mathbf{S_{sc}}$ and  
$\mathbf{S_{ss}}$). 
We have implemented this method and verified for some examples that it 
indeed provides the same results as the Arnoldi method. Actually it may 
even be more efficient than the direct Arnoldi method which may require a 
quite large Arnoldi dimension for a reliable first eigenvector. 
However, at this stage this approach also suffers from the same problem 
concerning the numerical inaccuracy for the cases of 
a very small core space gap. 
\begin{figure}[t]
\begin{indented}\item[]
\begin{center}  
\includegraphics[width=0.7\textwidth]{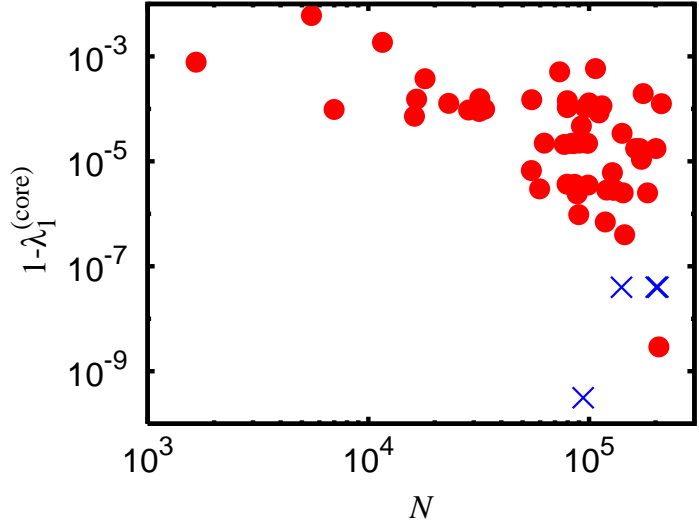}
% \vglue -0.3cm
\caption{Core space eigenvalue gap $1-\lambda_1^{\rm (core)}$ versus 
network size $N$ for the universities 
Glasgow, Cambridge, Oxford, Edinburgh, 
UCL, Manchester, Leeds, Bristol and Birkbeck (years 2002 to 2006) 
and Bath, Hull, Keele, Kent, Nottingham, Aberdeen, 
Sussex, Birmingham, East Anglia, Cardiff, York (year 
2006). Red dots correspond to data with 
$1-\lambda_1^{\rm (core)}>10^{-9}$ and blue crosses (shifted up 
by a factor of $10^9$) to the 
cases Cambridge 2002, 2003 and 2005 and Leeds 2006 with 
$1-\lambda_1^{\rm (core)}<10^{-16}$ where the maximal core space eigenvalue 
is determined by the projected power method. 
The data point at $1-\lambda_1^{\rm (core)}= 2.91\cdot 10^{-9}$ 
is for Cambridge 2004.
}
\label{fig4}
\end{center}
\end{indented}
\end{figure}

Fortunately the approach can be modified to be more accurate. To see this we 
use Eq.~(\ref{Smat_block}) and the fact that the columns of $\mathbf{S}$ 
are sum normalized which implies
$\|\mathbf{S_{sc}}\,\psi_1^{\rm (core)}\|_1+
\|\mathbf{S_{cc}}\,\psi_1^{\rm (core)}\|_1=1$ and therefore
\begin{equation}
\label{good_lambda_gap}
1-\lambda_1^{\rm (core)}=\|\mathbf{S_{sc}}\,\psi_1^{\rm (core)}\|_1
=\sum_{j\in V_{\rm SP}}\,\sum_{l\in V_c}\,S_{jl}\,\psi_{1}^{\rm (core)}(l)
\end{equation}
where $V_{\rm SP}$ denotes the set of subspace nodes and $V_c$ is the set 
of core space nodes (note that $\psi_{1}^{\rm (core)}(l)\ge 0$).
This expression, which relates the core space gap to 
the sum of all transitions from a core space node to a subspace node (the 
``escape probability'' from the core space), is the key to determine the 
gap accurately. 

First, we note that a numerically small core space gap (below $10^{-14}$) 
implies that 
the eigenvectors components $\psi_{1}^{\rm (core)}(l)$ are also 
numerically small for the core space nodes $l$ which are directly connected 
to a subspace node $j$ by a non-vanishing matrix element $S_{jl}>0$. 
To be more precise it turns out that for this situation 
the eigenvector $\psi_1^{\rm (core)}$ is strongly localized on a modest 
number of about $100$ nodes out of $10^5$ nodes in total and 
numerically small on the other nodes. Obviously, the nodes inside 
the small localization domain are not directly connected to a subspace node 
(by the matrix $\mathbf{S}$). The important point is that we can determine 
the eigenvector accurately also for the very small tails 
(below $10^{-15}$) by 
the {\em pure} power method (without intermediate Arnoldi steps) if we choose 
as initial vector a vector localized at the maximum node. The reason is that 
the non-vanishing matrix elements $S_{jl}$ connect only sites 
for which the eigenvector components are comparable 
in the order of magnitude. Therefore numerical round-off errors are minimized 
despite the fact that the resulting vector will contain components 
with a size ratio significantly above $10^{15}$ between maximal and minimal 
components. This is similar to certain localization problems in disordered 
quantum systems where it is in certain cases possible to determine numerically 
exponentially small tails of localized eigenvectors even if these tails are 
far below $10^{-15}$. 

Therefore, in practice, we  implement the following 
projected power method:
\begin{enumerate}
\item Determine a first approximation of $\psi_1^{\rm (core)}$ by the 
direct Arnoldi method which is accurate inside the localization domain 
but numerically incorrect for the very small tails on the nodes 
outside the localization domain. From these data we determine 
the node $l_{\rm max}$ 
at which $\psi_{1}^{\rm (core)}(l_{\rm max})$ is maximal. 
\item Choose as initial vector (on the full space including core space 
{\em and} subspace nodes) 
the vector localized on the node $l_{\rm max}$, 
i.e. $\psi(l)=\delta_{l,l_{\rm max}}$. 
\item \label{start} Make a copy of the vector: $\psi_{\rm old}=\psi$. 
\item Apply the matrix $S$ to the actual vector: $\psi=\mathbf{S}\,\psi$ which 
produces artificially non-zero values $\psi(j)$ on certain subspace nodes $j$.
\item According to Eq.~(\ref{good_lambda_gap}) 
compute the quantity $\sum_{j\in V_{\rm SP}}\,\psi(j)$ as 
approximation of the gap $1-\lambda_1^{\rm (core)}$.
\item Project the vector on the core space: $\psi(j)=0$ for all subspace 
nodes $j\in V_{\rm SP}$.
\item Normalize the vector by the 1-norm: $\psi=\psi/\|\psi\|_1$. 
\item Stop the iteration if $\|\psi-\psi_{\rm old}\|_1<\varepsilon_1$ 
and $\max_{\ l\in V_c}\,|\psi(l)-\psi_{\rm old}(l)|/|\psi(l)|<\varepsilon_2$. 
Otherwise go back to step \ref{start}.
\end{enumerate}
This algorithm produces an accurate vector very rapidly 
on the localization domain (less than 100 iterations) but in order to 
obtain an accurate value of the gap 
by Eq.~(\ref{good_lambda_gap}) the eigenvector needs to be accurate with 
a small relative error also in the very small tails and therefore the 
convergence criterion has to take into account the relative error for 
each component. We have chosen $\varepsilon_1=10^{-13}$ and 
$\varepsilon_2=10^{-6}$ which provides convergence with $10^6$ 
iterations for the cases of Cambridge 2002, 2003 and 2005. 
In the case of Leeds 2006 we even obtain convergence with 
$\varepsilon_1=\varepsilon_2=10^{-15}$ after $2\cdot 10^5$ iterations. 
For the particular case of Cambridge 2004 (where the gap $\sim 10^{-9}$ is 
still ``accessible'' by the Arnoldi method) the convergence is more difficult 
and we have stopped the iteration at $\varepsilon_1=10^{-12}$ 
and $\varepsilon_2=3.2\cdot 10^{-6}$. 

The choice of the initial vector localized at the maximum node is very 
important for the speed of the convergence. If we choose the delocalized 
vector $e/N$ as initial vector, it is virtually impossible to obtain 
convergence in the tails which stay at ``large'' values $\sim 10^{-8}$ 
unless we use intermediate Arnoldi steps but this destroys the fine 
structure of the tails below $10^{-15}$ which is crucial to 
determine the very small gap.

Using the above algorithm we obtain the gap values given in 
Table 2. 
\bigskip

\begin{table}[h]%
\small
\caption{\label{table2} Gap values}
\begin{tabular}{|l|c|}
\hline
 & $\ 1-\lambda_1^{\rm (core)}\ $ \\
\hline
\hline
\ Cambridge 2002\ \  & $\ 3.996\cdot 10^{-17}\ $ \\
\hline
\ Cambridge 2003\ \  & $\ 4.01\cdot 10^{-17}\ $ \\
\hline
\ Cambridge 2004\ \  & $\ 2.91\cdot 10^{-9}\ $ \\
\hline
\ Cambridge 2005\ \  & $\ 4.01\cdot 10^{-17}\ $ \\
\hline
\ Leeds 2006\ \  & $\ 3.126\cdot 10^{-19}\ $ \\
\hline
\end{tabular}
\bigskip
\end{table}

In Fig.~\ref{fig4} we compare these gap values to the other university 
networks for which we found by the Arnoldi method 
larger gaps $1-\lambda_1^{\rm (core)}>10^{-7}$. 

\begin{figure}[t]
\begin{indented}\item[]
\begin{center}  
\includegraphics[width=0.7\textwidth]{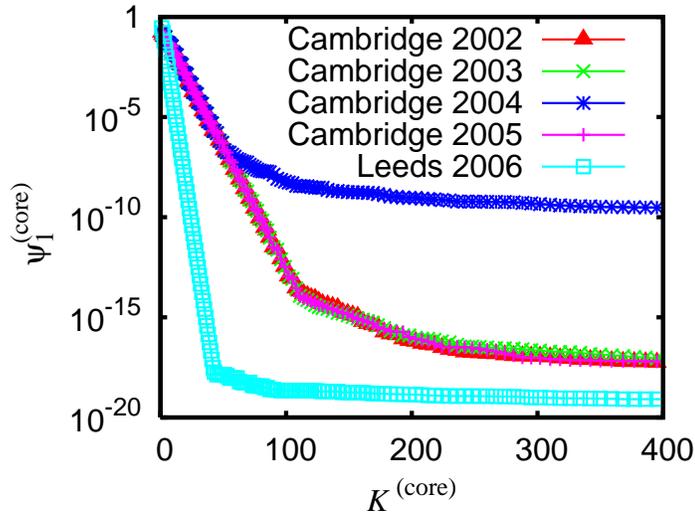}
% \vglue -0.3cm
\caption{First core space eigenvector $\psi_1^{\rm (core)}$ 
versus its rank index $K^{\rm (core)}$ for the university 
networks with a small core space gap $1-\lambda_1^{\rm (core)}<10^{-8}$.
}
\label{fig5}
\end{center}
\end{indented}
\end{figure}

In Fig.~\ref{fig5} we show the eigenvectors $\psi_1^{\rm (core)}$ 
obtained by the projected power method versus their rank index 
$K^{\rm (core)}$ defined by the ordering of the components of theses vectors. 
We can clearly identify the exponential localization on 40 nodes 
for Leeds 2006 or 110 nodes for Cambridge 2002, 2003 and 2005 with values 
below $10^{-18}$ (Leeds 2006) or $10^{-14}$ (Cambridge 2002, 2003 and 2005). 
The case Cambridge 2004 with a quite larger gap $\sim 10^{-9}$ 
provides at first the same exponential localization as the other three 
cases of Cambridge but after 50 nodes it goes over to a tail in the 
range $10^{-8}$ to $10^{-10}$. In all cases the range of values of the 
small tail 
is in qualitative agreement with the gap values in the Table 2 and 
the expression (\ref{good_lambda_gap}). 

When the iteration with the matrix $\mathbf{S}$ starts at the maximal node 
the vector diffuses first quite slowly inside the localization domain 
for a considerable number of iterations (46 
for Leeds 2006 and 35 for Cambridge 2002, 2003 and 2005) until it reaches 
a dangling node at which point the diffusion immediately extends to the 
full network since the dangling node is artificially connected to all nodes. 
However, at this point the probability of the amplitude is already 
extremely small. Therefore the initial node belongs technically to the 
core space (since it is ``connected'' to all other nodes) but practically 
it defines a quasi subspace (since the probability to leave the 
localization domain is very small $\sim 10^{-19}$ or $\sim 10^{-17}$). At 
$1-\alpha=10^{-8}$, which is much larger than the gap, this quasi subspace 
also contributes to the PageRank in the same way as the exact invariant
subspaces. 
This provides somehow a slight increase of the effective value of $N_s$ 
but it does not change the overall picture as described in Section 2. 

Fig.~\ref{fig5} also shows that apparently the particular network structure 
responsible for this quasi subspace behavior is identical for the three 
cases Cambridge 2002, 2003 and 2005. 
For Cambridge 2004 this structure also exists but here there is one 
additional dangling node which is reached at an earlier point of the 
initial slow diffusion providing delocalization on a scale 
$\sim 10^{-10}-10^{-8}$. 
For the case of Cambridge 2006 with a ``large'' gap $\sim 10^{-4}$ 
this structure seems to be completely destroyed but this may be due 
to one single modified matrix element $S_{jl}$ if compared to the networks 
of the previous years. 

\section{Universal properties of PageRank and subspace distribution}

\begin{figure}[t]
\begin{indented}\item[]
\begin{center}  
\includegraphics[width=0.7\textwidth]{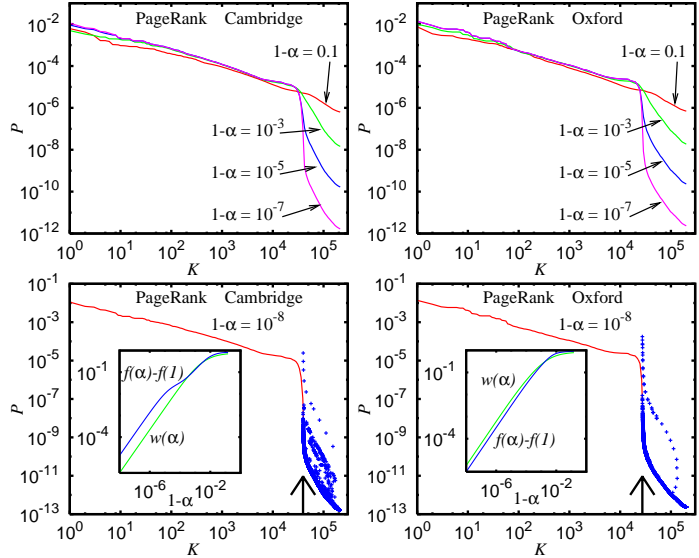}
\caption{Left panels (right panels) correspond to Cambridge 2006 (Oxford 2006).
{\em Top row:} PageRank $P(K)$  for 
$1-\alpha=0.1,\,10^{-3},\,10^{-5},\,10^{-7}$. 
Numerical precision  is such that $\| P-G(\alpha)P\|_1\,<10^{-13}$. 
{\em Bottom row:} $P(K)$  at $1-\alpha=10^{-8}$.
Blue crosses correspond to the eigenvector of the largest core space 
eigenvalue $\lambda^{\rm (core)}_1=0.999874353718$ (0.999982435081) 
multiplied by $(1-\alpha)/(1-\lambda^{\rm (core)}_1)$. 
The arrow indicates 
the first position where a site of the core space $V_c$ contributes to 
the rank index; all sites at its left are in an
invariant subspace. Insert shows the residual weight $w(\alpha)$
with $w(\alpha)=\sum_{j\in V_c} P^{(\alpha)}(j)$ of the core space $V_c$ 
in the PageRank 
and the difference $f(\alpha)-f(1)$ versus $1-\alpha$ where $f(\alpha)$ is 
the PageRank fidelity with respect to $\alpha=0.85$, i.e. 
$f(\alpha)=\<P^{(\alpha)}\,|\,P^{(0.85)}\>/(\|P^{(\alpha)}\|_2\,\|P^{(0.85)}\|_2)$. 
Note that $\|P^{(\alpha)}\|_2\neq 1$ since the PageRank is normalized 
through the 1-norm: $\|P^{(\alpha)}\|_1=1$. 
The limiting value $f(1)=0.188400463202$ (0.097481331613) is 
obtained from linear extrapolation from the data with smallest values of 
$1-\alpha$ which we verified to be exact up to machine precison. 
}
\label{fig6}
\end{center}
\end{indented}
\end{figure}

Using the powerful numerical methods described above we turn to 
the analysis of universal properties of  PageRank. 
Fig.~\ref{fig6} clearly confirms the theoretical picture 
given in section 2 of the limit behavior for the 
PageRank at  $\alpha \rightarrow 1$. In particular one can clearly 
identify the limit where it is localized in 
the invariant subspaces 
\cite{footnote} 
with only
small corrections  $\sim(1-\alpha)$ at the core space nodes. We also 
determine the eigenvector of the largest core space eigenvalue 
$\lambda_1^{\rm (core)}$ of the projected matrix $\mathbf{S_{cc}}$. In the 
lower panels of Fig. \ref{fig6}, we compare the PageRank at 
$1-\alpha=10^{-8}$ with this vector (normalized by the 1-norm) multiplied by 
$(1-\alpha)/(1-\lambda_1^{\rm (core)})$. We observe that except for 
a very small number of particular nodes this vector approximates quite well 
the core space correction of the PageRank even though the corrections 
due to the second term in (\ref{PageRank2}) are more complicated 
with contributions from many eigenvectors. In the inserts, we also show 
the fidelity of the PageRank, which decays from 
1 at $1-\alpha=0.15$ to about 0.188 (0.097) at $1-\alpha=10^{-8}$, and 
the residual weight 
$w(\alpha)=\sum_{j\in V_c} P^{(\alpha)}(j)$ of the core space $V_c$ 
in the PageRank which behaves as $w(\alpha)\approx 221.12\,(1-\alpha)$ 
[$\approx 607.12\,(1-\alpha)$] for $1-\alpha<10^{-5}$.

\begin{figure}[t]
\begin{indented}\item[]
\begin{center}  
\includegraphics[width=0.7\textwidth]{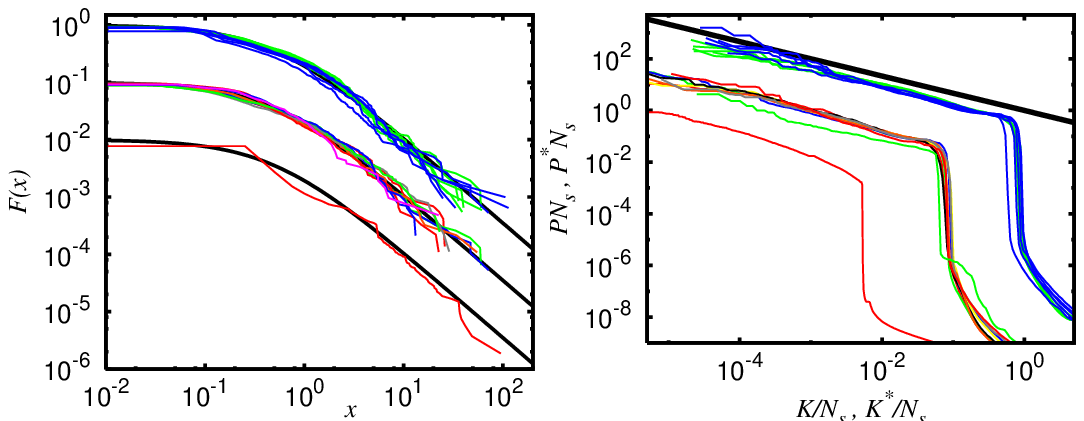}
% \vglue -0.3cm
\caption{{\em Left panel:} 
Fraction of invariant subspaces $F$ with dimensions larger than $d$ 
as a function of the rescaled variable $x=d/\langle d\rangle$. Upper
curves correspond to Cambridge (green) and Oxford (blue) for years 
2002 to 2006 and middle curves (shifted down by a factor of 10) 
to the university networks of Glasgow, Cambridge, Oxford, Edinburgh, 
UCL, Manchester, Leeds, Bristol and Birkbeck for year 2006 
with $\langle d\rangle$ between 14 and 31.
Lower curve (shifted down by a factor of 100) corresponds to the 
matrix $\mathbf{S^*}$ of Wikipedia with $\langle d\rangle=4$. 
The thick black line is 
$F(x)=(1+2x)^{-1.5}$. 
{\em Right panel:} Rescaled PageRank $P\,N_s$ versus rescaled 
rank index $K/N_s$ for $1-\alpha=10^{-8}$ and 
$3974 \leq N_s \leq 48239$ 
for the same university networks as in the left panel 
(upper and middle curves, the latter shifted down and left by a factor of 10). 
The lower curve (shifted down and left by a factor of 100) shows the rescaled 
CheiRank of Wikipedia $P^*\,N_s$ versus $K^*/N_s$ with $N_s=21198$. 
The thick black line 
corresponds to a power law with exponent $-2/3$.
}
\label{fig7}
\end{center}
\end{indented}
\end{figure}

As mentioned in the previous Section,
we also determine the subspace structure and the PageRank at 
$1-\alpha=10^{-8}$ for other university networks available at
\cite{uniuk} and for the matrix $\mathbf{S^*}$ of Wikipedia 
\cite{wiki} with $N=3282257$ and
$N_s=21198$ (it turns out that the matrix 
$\mathbf{S}$ for Wikipedia provides only very few small size subspaces with 
no reliable statistics). 
A striking feature is that the distribution of subspace  dimensions 
$d_j$ is universal for all networks considered 
(Fig.~\ref{fig7} left panel). The fraction of 
subspaces with dimensions larger than $d$ is well described by the 
power law $F(x)=(1+x/(b-1))^{-b}$ 
with the dimensionless variable 
$x=d/\langle d\rangle$, where
$\langle d\rangle$ is the average 
subspace dimension. The fit of all cases gives
$b= 1.608 \pm 0.009 \approx 1.5$. It is interesting to note
that the value of $b$ is close to the exponent of
Poincar\'e recurrences in dynamical systems \cite{ulamfrahm}.
Possible links with 
the percolation on directed networks 
(see e.g. \cite{schwartz}) are still to be elucidated.
The rescaled Pagerank $PN_s$ (or CheiRank $P^*N_s$ for the 
case of Wikipedia) takes a universal form with a power law 
$P\sim K^{-c}$ for $K<N_s$ with an exponent  
$c= 0.698 \pm 0.005 \approx 1/b =2/3$ and $P\sim (1-\alpha)$ 
close to zero for $K>N_s$ (see right panel of Fig.~\ref{fig7}).

For certain university networks, Cambridge 
2002, 2003 and 2005 and Leeds 2006, 
there is a specific complication. Indeed, the AM 
(with $n_A=10000$) provides a maximal 
core space eigenvalue $\lambda_1^{\rm (core)}$ 
{\em numerically} equal to 1, which should not be possible. 
A more careful evaluation by a different algorithm, 
based on the power method (iterating $\mathbf{S}$ with a subsequent core 
space projection) and measuring the 
loss of probability at each iteration, shows that this eigenvalue is
indeed very close but still {\em smaller} than 1. For the three cases 
of Cambridge we find 
$1-\lambda_1^{\rm (core)}\approx 4.0\cdot 10^{-17}$ and for Leeds 2006:
$1-\lambda_1^{\rm (core)}\approx 3.1\cdot 10^{-20}$
(see details in Section 5). The corresponding 
eigenvectors are exponentially localized on a small number of nodes 
(about 110 nodes for Cambridge and 40 nodes for Leeds 2006) being very small 
($<10^{-14}$ for Cambridge and $<10^{-18}$ for Leeds 2006) on other 
nodes.  These quasi-subspaces with 
small number of nodes belong {\em technically} 
to the core space, since they are eventually linked to a dangling node, but 
when starting from the maximal node of these eigenvectors it takes a 
considerable number of iterations with a strong reduction of probability 
to reach the dangling node. Since their eigenvalue is very close to 1, 
these quasi-subspaces also contribute to the PageRank at $1-\alpha=10^{-8}$ 
in the same way as the exact invariant subspaces.
However, since the size of these 
quasi-subspaces is  small they do not change 
the overall picture and we can still identify a region of large PageRank 
with $N_s$ subspace or quasi-subspace nodes and vanishing PageRank for 
the other core space nodes. For most of the other universities and also the 
matrix $\mathbf{S^*}$ of Wikipedia we have $1-\lambda_1^{\rm (core)}\ge 10^{-6}$ 
(and $1-\lambda_1^{\rm (core)}\sim 10^{-9}$ for Cambridge 2004).

\section{Discussion}

Our results show that for $\alpha \rightarrow 1$ the PageRank vector 
converges  to a universal distribution $P \sim 1/K^c$ 
determined by the invariant subspaces (with $c \approx 2/3$).
The fraction of
nodes which belong to these subspaces varies greatly depending on the
network, but the distribution of the subspace sizes
is described by a universal function
$F(x)=1/(1+2x)^{3/2}$ that reminds the properties of critical
percolation clusters. When $\alpha$ decreases from $1$, the PageRank
undergoes a transition which allows to properly rank all nodes. 
This process is controlled by the largest eigenvalues of the
core matrix $\mathbf{S_{cc}}$, which are strictly below $1$
but can be extremely close to it. 
Their distance from
$1$ sets the scale of the transition, and the associated eigenvectors
of $\mathbf{S_{cc}}$ control the new ranking of nodes. Although 
at $\alpha=1$ the eigenspace for eigenvalue $1$ can be very large, 
for $\alpha$ sufficiently larger
in norm than the eigenvalues of $\mathbf{S_{cc}}$, 
the PageRank remains fixed when
$\alpha \rightarrow 1$, in a way reminiscent of degenerate perturbation 
theory in quantum mechanics.  
Our highly accurate numerical method based on alternations 
of Arnoldi iterations and direct iterations of $\mathbf{G}$ 
matrix enables to determine the correct PageRank even 
where the scale of this transition
is extremely small ($1-\lambda_1^{\rm{(core)}} \approx 10^{-20}$) 
and the matrix size is very
large (up to several millions).  The very slow convergence of
the power method in this regime is reminiscent of very long equilibration
times in certain physical systems (e.g. spin glasses), 
and thus  Arnoldi iterations  can be viewed as a certain kind of simulated 
annealing process which
enables to select the correct eigenvector among
many others with very close eigenvalues.  The PageRank in this regime of
$\alpha \rightarrow 1$ shows universal properties being different
from the usual PageRank
at $\alpha \approx 0.85$, with a different statistical distribution.
This can be used to refine search and ranking in complex networks
and hidden communities extraction.

Finally we note that usually in quantum physics one deals with unitary matrices
with a real spectrum. In the case of directed Markov chains we naturally
obtain a complex spectrum.  In physical quantum systems 
a complex spectrum appears in positive quantum maps \cite{karol1},
problems of decoherence and quantum measurements \cite{karol2} and
random matrix theory of quantum chaotic scattering \cite{guhr}.
Thus we hope that a cross-fertilization between complex matrices and
directed network will highlight in a new way 
the properties of complex networks.

\ack
% {\bf Acknowledgments:}
We thank CalMiP for supercomputer access and A.D.Chepelianskii
for help in data collection from \cite{uniuk}.

\normalsize
%\newpage


\begin{thebibliography}{99}
\bibitem{footnote} In certain invariant subspaces, there are nodes
        with no ingoing links from the same subspace, which do not 
         contribute 
        to the PageRank for $\alpha\rightarrow 1$. Except for 
        Wikipedia (CheiRank), they are very few in our data and 
        their effect is not visible in the figures.
\end{thebibliography}
\end{document}